# Assessment of Brightness Mitigation Practices for Starlink Satellites

Anthony Mallama[1*], Andreas Hornig[2], Richard E. Cole,
Scott Harrington, Jay Respler[1], Ron Lee and Aaron Worley

2023 October 1

\* Correspondence: anthony.mallama@gmail.com
[1] IAU - Centre for the Protection of Dark and Quiet Skies from Satellite Constellation Interference
[2] University of Stuttgart, Germany

## Abstract

Photometric characteristics for all models of Starlink satellites launched to date are reviewed. The Original design that lacked brightness mitigation is the most luminous. SpaceX installed a sunshade on the VisorSat model which reduced its luminosity by a factor of 3. The visor was omitted on Post-VisorSat spacecraft with laser communication which followed, but the company added a reflective layer which resulted in an intermediate brightness between Original and VisorSat. SpaceX is applying advanced brightness mitigation techniques to their Generation 2 Starlink satellites which are larger. The first of these, called Minis, are dimmer than Gen 1 Starlinks despite their greater size. Photometric observations verify that brightness mitigation efforts employed by SpaceX reduce spacecraft luminosity substantially. However, the satellites still have some negative impact on astronomical observations and the very large satellites planned for later in Gen 2 may interfere more seriously.

## 1. Introduction

Satellite constellations are beginning to impact the work of professional astronomers as reported by Barentine et al. (2023). They point out that space objects leave streaks on images which can reduce their scientific potential. Additionally, smaller objects elevate the diffuse brightness of the sky. The authors compute the potential increase in sky brightness and address the corresponding loss of astronomical information.

Amateur astronomers and others who appreciate the aesthetics and cultural significance of the night sky are also adversely affected by satellites as discussed by Mallama and Young (2021). Spacecraft brighter than magnitude 6 are distractions visible to the unaided eye, while those brighter than 7 impact professional research.

SpaceX operates the largest satellite constellation with more than 4,000 Starlink spacecraft already in orbit and regulatory approval for many more. The initial launch of 60 satellites on one rocket in 2019 raised concerns because of their brightness. SpaceX responded by making several changes to the spacecrafts' physical designs and to their satellite operations. This paper reviews the brightness mitigation strategies and the corresponding luminosity changes recorded by observers.

Section 2 defines the terminology used in this paper. Section 3 summarizes the brightness mitigation techniques implemented by SpaceX. Section 4 describes the methods of photometry used to record satellite magnitudes. Section 5 characterizes the luminosity of Starlink satellites as derived from observed magnitudes. Section 6 describes numerical modeling of spacecraft brightness and illustrates how the models fit photometric observations. Section 7 discusses the impact of Starlink satellites on astronomy and addresses international efforts to mitigate the negative effects of all satellite constellations. Our conclusions are given in Section 8.

## 2. Definitions and abbreviations

The terms elevation, height and range are differentiated as follows in this paper. Elevation is the angular distance of a satellite above the Earth's horizon measured in degrees. Height refers to the vertical distance of a satellite above the Earth's surface in km. Range is the distance between an observer and a spacecraft in km. The term altitude is not used here to avoid confusion.

The observed brightness of a satellite is its apparent magnitude. That luminosity may be adjusted to a standard distance of 1000 km by applying the inverse square law of light. The distance-adjusted brightness, or 1000-km magnitude in this paper, is useful for comparing satellite luminosities measured at different ranges. Magnitudes may also be adjusted to 550 km which was the orbital height of early Starlink satellites. The 550-km values are referred to as characteristic magnitudes because they correspond to the brightness of many Starlink satellites when they are near the observer's zenith.

Statistical means sometimes include variational parameters. The standard deviations, abbreviated as SD, represent the scatter about the mean. The standard deviation of the mean, SDM, is its formal uncertainty.

A bidirectional reflectance function defines how light is reflected from a surface. The BRDF is used in conjunction with the physical layout of a satellite's component parts. In the case of Starlink spacecraft, the main components are its antenna panel and solar array as shown in Figure 1. Parameters of the BRDF model may be adjusted to fit observed magnitudes.

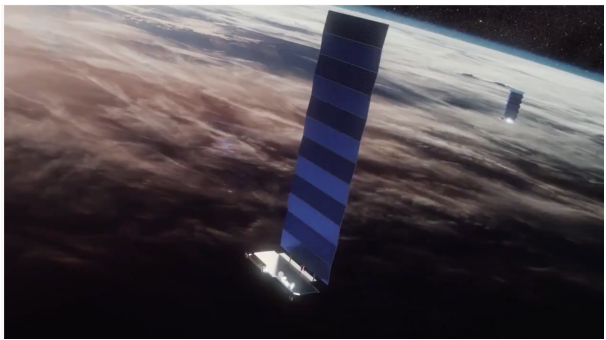

*Fig. 1. The horizontal component of Starlink is the antenna panel and the vertical one is the solar array. Illustration from SpaceX.*

Phase angle is the arc measured at the satellite between directions to the Sun and to the observer. This angle is used to characterize satellite brightness and it leads to the phase function which is brightness as the dependent variable of phase angle.

Orbit-raise is the phrase used by SpaceX in referring to satellites that are ascending from their injection heights to higher orbits. Parking orbits are where low height satellites wait for precession to change their orbital plane. On-station satellites are those which have attained their final heights. Spacecraft attitude refers to the orientation of the satellite in space especially with respect to the Sun and the observer. Lastly, SpaceX uses the term conops to mean 'concept of operations'.

### 3. Brightness mitigation practices

This Section reviews the strategies employed by SpaceX to dim Starlink satellites. The corresponding changes of observed brightness are also mentioned qualitatively. Quantitative photometry is addressed in Sections 4 and 5.

The Original model of Starlink spacecraft consisted of an antenna panel measuring 1.3 x 2.8 m and a solar array 2.8 x 8.1 m, with a total surface area of 26.32 m$^2$. These dimensions remained unchanged until the second generation of spacecraft described later in this Section.

No brightness mitigation measures were implemented for the Original satellites because their impact on astronomy was not foreseen. In 2020 SpaceX applied a low albedo coating to a test satellite named DarkSat. Tregloan-Reed et al. (2020) and Halferty et al. (2022) found it to be dimmer but Takahashi et al. (2020) reported that it was brighter. In any case, the spacecraft absorbed too much sunlight which caused thermal problems and this approach was abandoned.

The next design change was incorporated into the VisorSat model of Starlink. The 'visor' refers to a shade that prevents sunlight from reaching the underside of the antenna panel which faces observers on the ground. This modification reduced the brightness of satellites on-station substantially (Mallama 2021a and 2021b, Krantz et al. 2023 and Halferty et al. 2022). However, SpaceX stopped attaching visors on the next model of Starlink satellites which used laser communication because they interfered with the beam.

The spacecraft model that followed VisorSat is referred to herein as Post-VisorSat. While these satellites lacked the Sun shade, SpaceX applied a dielectric reflective layer to the bottom of the antenna panel, as shown in Figure 2, which directed sunlight into space rather than allowing it to scatter toward the ground. The Post-VisorSat spacecraft on-station were found to be intermediate in brightness between Original and VisorSat satellites by Mallama and Respler (2022) and by Krantz et al. (2023).

Additionally, SpaceX changed the roll angle for VisorSat and Post-VisorSat spacecraft in order to mitigate their brightness. This 'knife-edge' attitude, which was applied to satellites in orbit-raising, placed the Sun in the plane of their flat surfaces. Mallama and Respler (2023) found that knife-edge configuration reduced luminosity in the early mission phases.

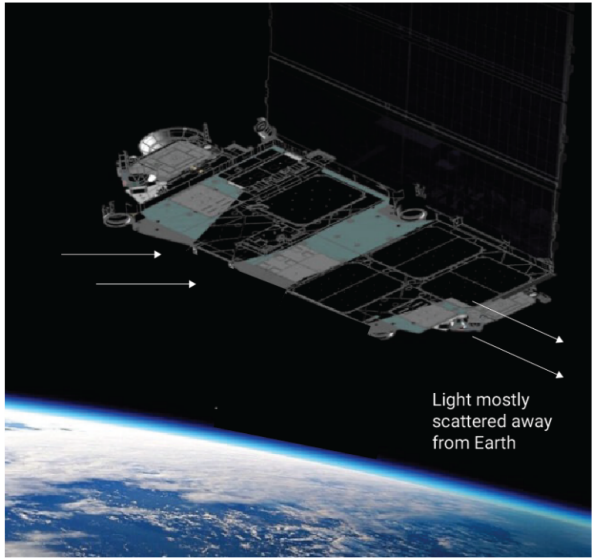

*Fig. 2. Reflective surfaces direct sunlight away from observers on the ground. Illustration from SpaceX.*

SpaceX began launching their second-generation Starlink spacecraft in 2023. The first model is called Mini because it is smaller than the full-sized Gen 2 satellites which will follow. The antenna panels of Mini satellites measure 2.7 x 4.1 m and their two solar panels are each 4.1 x 12.8 m. The total surface area of 116.0 $m^2$ is more than four times that of Gen 1 spacecraft.

Surface area usually correlates with brightness. So, astronomers were especially concerned about the luminosity of Gen 2 spacecraft. However, SpaceX made two changes to reduce the brightness of these satellites. First, they improved the mirror-like reflective layer on the antenna panel so that more sunlight is directed into space. Second, they developed a conops similar to knife-edge and implemented it for on-station satellites. This configuration points the plane of the solar arrays toward the Earth's limb when the satellites are near the terminator. Thus, observers only see their dark sides as shown in Figure 3. Mallama et al. (2023) found that the mitigation strategy is effective in reducing the brightness of Mini satellites.

This Section has addressed brightness mitigation strategies implemented by SpaceX. The next Section describes the methods used to measure Starlink satellite magnitudes. In Section 5 we examine the observational results for each spacecraft model more thoroughly.

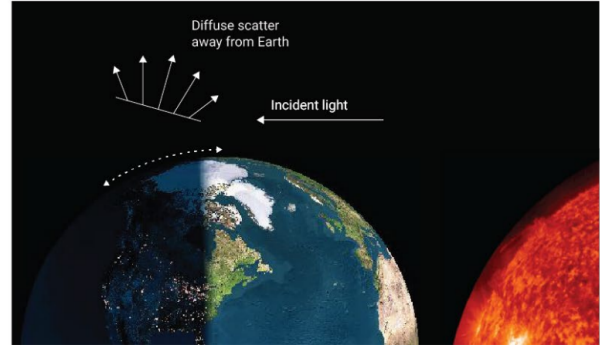

*Fig. 3. Observers only see the dark side of solar arrays. Illustration from SpaceX.*

**4. Observation methods**

Starlink brightness measurements have been acquired by several different techniques. These include visual perception by the human eye, recordings made with a digital camera used in video mode, output from a wide-field 9 channel system with sCMOS sensors, and telescopic observations recorded by solid state sensors.

Visual observers record Starlink magnitudes by comparing their brightness to nearby reference stars. Angular proximity between the spacecraft and those stellar objects accounts for variations in sky transparency and sky brightness. The perceptual method of observing is described more thoroughly by Mallama (2022a).

Video observations were recorded with a Sony Alpha A7s-I camera and a Sony FE 1.8/50 lens. The Astrometry.net application was run on a Raspberry Pi 4 device for extracting information about the stars. Specially coded python software was executed on a Windows computer to perform the overall measurements and data processing. Magnitudes from video frames were averaged over five second time intervals to form a mean value. This system is the prototype optical ground station (OGS) for the Distributed Ground Station Network (DGSN) being developed at the University of Stuttgart. The DGSN project was started within the SmallSat-Design-Studies at the Institute of Space Systems (IRS). It was part of several annual Google and ESA Summer of Code campaigns. The DSGN is a PhD-research topic at the Institute for Photogrammetry (IFP) at the University of Stuttgart.

Observations were also gathered from the database of the MMT9 system described by Karpov et al. (2015) and Beskin et al. (2017). This robotic observatory consists of nine 71 mm diameter f/1.2 lenses and 2160 x 2560 sCMOS sensors. The detectors are sensitive to the visible spectrum from red through blue. We collected their apparent magnitudes along with date/time values and computed other quantities needed for analysis.

The methods described above were utilized by the authors of this paper to obtain observational data, and magnitudes collected from the MMT9 database have also been used in our studies. The magnitude scales for all these techniques closely agree. MMT9 values are within 0.1 magnitude of the V-band based on information in a private communication from S. Karpov as discussed by Mallama (2021). The video magnitudes match visual and V-band results closely because the camera is panchromatic in visible light. That agreement is shown empirically by Mallama et. al. (2023).

Additional observations have been reported by other groups. Their instruments include the Pomenis LEO Satellite Photometric Survey Telescope at Mt. Lemmon in Arizona USA (Krantz et al. 2023), the Chakana 0.6-m telescope in Chile (Tregloan-Reed et al. 2020), the Stingray prototype consisting of a telephoto lens and CMOS sensor also located in Arizona (Halferty et al. 2022), the Zwicky Transit Facility which uses the Schmidt telescope at Palomar Observatory (Mroz et al. 2022), the Plaskett 1.6 m telescope of the Dominion Astrophysical Observatory (Boley et al. 2022), the SCUDO telescope in Italy (Hossein et al. 2022) and an ensemble of eight different telescopes (Takahashi et al. 2023).

**5. Empirical brightness characterization**

This Section characterizes the brightness of all four models of Starlink satellites that have been launched to date. Mean magnitudes, phase functions and brightness surges are discussed.

*5.1 Original design is brightest*

The first photometric survey of Starlink satellites was performed by McDowell (2020) using visual magnitudes from the [SeeSat](#) email archive. He found that magnitudes spanned 'from 3 to 7 with most between visual mag $5.5 \pm 0.5$' for satellites on-station at 550 km.

A follow-up study combining visual magnitudes from SeeSat with V-band magnitudes from MMT9 was conducted by Mallama (2020). The 830 luminosities for on-station satellites were adjusted to the standard 1000-km distance. The mean of adjusted magnitudes was 5.93 +/-0.67 +/-0.02, where the first variational quantity is the SD and the second is the SDM. When the mean 1000-km magnitude is re-adjusted to 550 km, which is the height of those on-stations spacecraft, the characteristic magnitude is 4.63.

Mallama also reported on brightness surges or 'flares'. Very bright flares spanning from magnitude -3 to -8 were reported on 8 occasions for orbit-raising satellites between 380 and 425 km. The Original design of Starlink satellites is still the brightest of all models in terms of their mean magnitudes and their flares.

*5.2 VisorSat is Fainter than Original*

SpaceX added a visor to this model in order to prevent sunlight from reaching the underside of the antenna panel which faces observers on the ground. Several studies quantified the effectiveness of this brightness mitigation.

Takahashi et al. (2023) recorded 19 observations of the first VisorSat spacecraft and 12 of Original design satellite Starlink-1113, each in 8 filters. They found that VisorSat was generally dimmer than the other spacecraft.

Halferty et al. (2022) recorded 363 GAIA G magnitudes of Original and VisorSat spacecraft. Their results indicate that the brightness mitigation applied to VisorSats dimmed them by an average of 0.9 magnitudes or a luminosity factor of 2.3.

Mallama (2021a) analyzed 430 visual and MMT9 magnitudes for on-station VisorSats. The mean of 1000-km mags was 7.22 +/- 0.85 +/- 0.04. Adjustment to the 550 km on-station height of these spacecraft indicated a characteristic mag of 5.92. The difference between these results and those for Original design (Mallama, 2020) is 1.29 magnitudes which corresponds to a factor of 3.2 in dimming.

In a large-scale study of MMT9 data Mallama (2021b) analyzed more than 60,000 VisorSat magnitudes and over 40,000 Original mags for on-station spacecraft. The mean of 1000-km magnitudes was 7.21 +/- 0.89 +/- 0.01 for VisorSats and 5.89 +/- 0.46 +/- 0.01 for Originals. The characteristic magnitudes at a distance of 550-km are 5.91 and 4.59. The difference of 1.32 magnitudes implies that VisorSats were dimmer by a factor of 3.3.

This study also compared the size and frequency of flare events of these two models. The light curve of a large flare is shown in Figure 4.

The data in Table 1 indicate that VisorSats produce more flares than Originals. The mean intervals between flares exceeding 0.5 magnitude were 129 seconds for VisorSats and 622 seconds for Originals. The percentage of the elapsed time spent above threshold amplitudes of 0.5, 1.0 and 2.0 magnitudes are also listed in the Table. They vary from 0.0% for flares of Original satellites exceeding 1.0 magnitude to 2.8% for VisorSat flares of 0.5 mag.

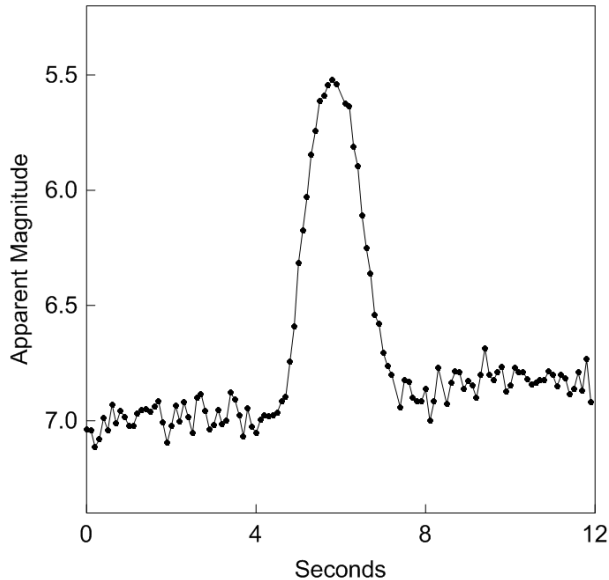

Fig. 4. A flare of Starlink-1538 recorded by MMT9 on 2021 May 2. Illustration from Mallama (2021b).

Table 1. Flare amplitude and frequency

|  | Mean Interval (seconds) | Time Percentage (at amplitude) | | |
|---|---|---|---|---|
|  |  | 0.5 | 1.0 | 2.0 |
| Original | 622 | 0.4 | 0.0 | 0.0 |
| Visorsat | 129 | 2.8 | 1.0 | 0.1 |

Finally, Hossein et al. (2022) obtained 571 RGB magnitudes for Original and VisorSat spacecraft. They analyzed the data as a function of satellite heights, ranges and other parameters. However, the results did not distinguish between Originals and VisorSats. So, no brightness comparison can be reported here.

*5.3 Post-VisorSats are intermediate in brightness*

When SpaceX added lasers to Starlink satellites they stopped including visors because these structures blocked the light beams. The omission of visors would have returned the brightness of Post-VisorSat spacecraft to approximately that of Originals. However, the company added a dielectric layer (Figure 2) to the bottom of the antenna panel for brightness mitigation. This mirror-like surface directed sunlight into space rather than allowing it to scatter toward observers on the ground.

Mallama (2022b) analyzed 58 visual magnitudes for on-station Post-VisorSats and 44 for VisorSats recorded by J. Respler in 2022. After adjustment for distance the Post-VisorSat spacecraft averaged 0.5 mags brighter than VisorSat. Nevertheless, they were 0.8 mags fainter than the Original design.

*5.4 Comparison of all three models from Generation 1*

Mallama and Respler (2022) analyzed a uniform set of visual magnitudes which they had recorded for on-station Original design, VisorSat and Post-VisorSat spacecraft. Figure 5 demonstrates that Original is the brightest followed by Post-VisorSat and VisorSat. A more recent set of video magnitudes for all three Gen 1 models, also shown in the figure, indicates the same ordering of brightness.

Krantz et al. (2023) reported findings similar to Mallama and Respler (2022). Their median apparent magnitudes for Original Design, VisorSat and Post-VisorSat are 5.72, 6.87 and 6.15, and the corresponding interdecile ranges span 2.58, 2.90 and 2.59 magnitudes, respectively. They point out that the brightness distribution is not statistical randomness.

An important aspect of the phase functions shown in Figure 5 is their concave upwards curvature. High luminosity at small phase angles is expected because the satellites are nearly opposite the Sun from the observer and so are almost fully lit. However, the brightness at large phase angles occurs when the spacecraft are between the Sun and the observer. In that case the high luminosity indicates forward scattering from back-lit components.

Krantz et al. (2023) reported excess brightness for satellites 'at mid-elevations opposite the Sun with an additional hot spot at low solar elongation above the below-horizon Sun'. These areas of the sky are equivalent to low and high phase angles, respectively. The great luminosity at high phase angles is due to satellites reflecting light from the dayside of the Earth. This phenomenon is discussed more fully in the next section which describes BRDF modeling.

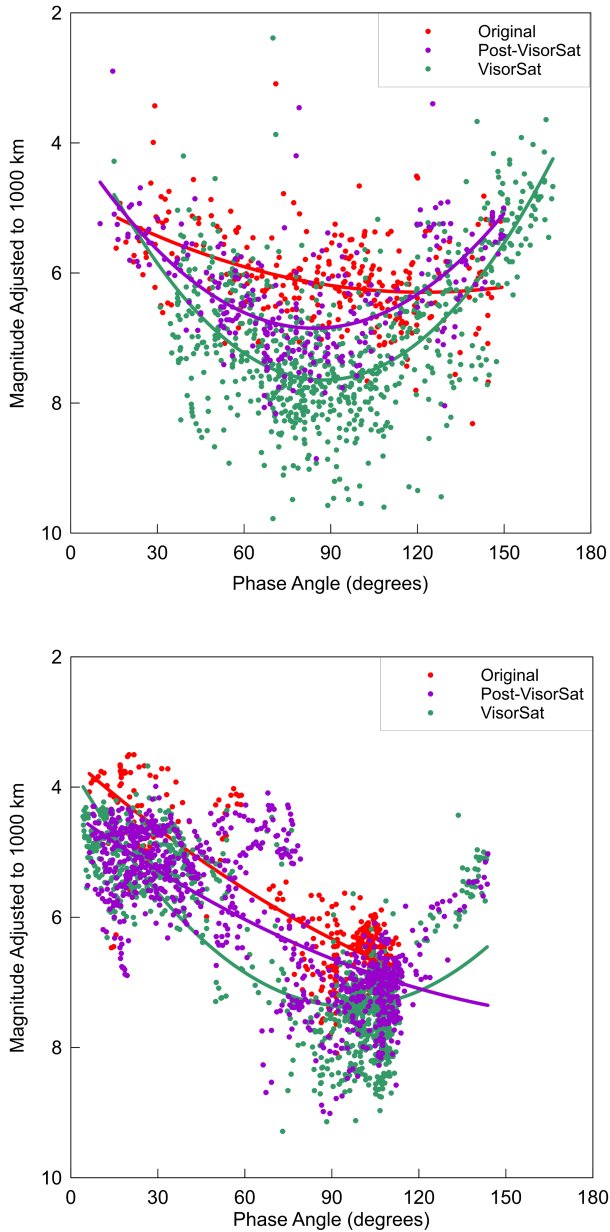

reduce distance-adjusted brightness by a factor of 10 as illustrated in Figure 6.

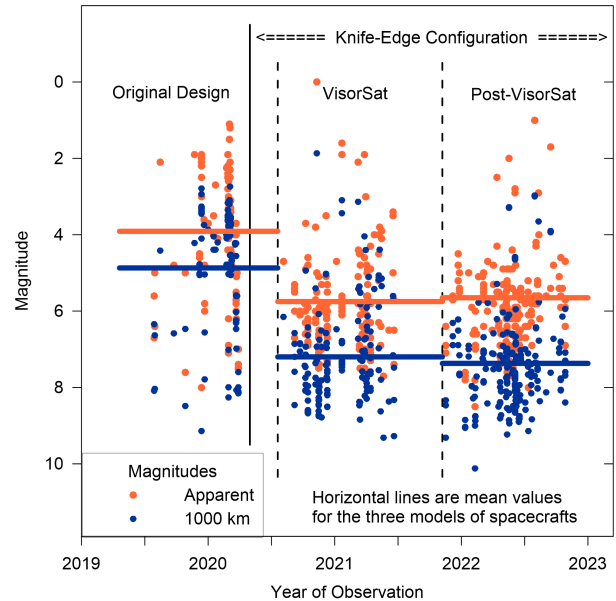

*Fig. 6. The knife-edge technique of brightness mitigation reduced luminosity for apparent and for distance-adjusted magnitudes. Illustration from Mallama and Respler (2023).*

*5.5 Gen 2 Mini satellites are largest and faintest*

Mini satellites have a surface area of 116 m$^2$ which is more than 4 times that of Gen 1 spacecraft. They are called 'Mini' because regular Gen 2 spacecraft will be even larger. The increased size concerned astronomers because bigger satellites are usually brighter. However, SpaceX instituted an aggressive strategy for brightness mitigation to compensate for the larger dimensions. They improved the dielectric layer on the bottom of the antenna panel which directed more sunlight back into space (Figure 2). They also developed a new conops, similar to knife-edge, for on-station spacecraft where the planes of the solar arrays point to the Earth's limb (Figure 3) when satellites are near the terminator. Observers on the ground only see the dark sides of the arrays in this attitude.

Mallama et al. (2023) found that mitigation reduced the brightness of on-station Mini satellites when compared to spacecraft observed during early mission phases without mitigation. The means of apparent magnitudes for mitigated spacecraft along with their SDs and SDMs were 7.06 +/- 0.91 +/- 0.10 and the values for magnitudes adjusted to 1000-km distance were 7.87 +/- 0.79 +/- 0.09. The corresponding statistics for satellites recorded during early mission phases were

*Fig. 5: Individual magnitudes and best-fit quadratic phase functions for the three models of Gen 1 Starlink satellites illustrate that Original design is brightest and VisorSat is faintest over most of the observed phase angles. Visual data are plotted in the panel on the top and video data are on the bottom. Subtract 1.3 magnitudes to adjust to 550-km.*

Mallama and Respler (2023) also examined the effectiveness of roll angle adjustment in dimming different models of Gen 1 satellites. SpaceX developed this knife-edge technique which places the Sun in the plane of flat surfaces on the satellites for brightness mitigation. The company applied it during orbit-raising to VisorSat and Post-VisorSat spacecraft but not in time for Originals. Roll angle adjustment was found to 3.97 +/- 1.96 +/- 0.09 and 5.08 +/- 1.70 +/- 0.08. The difference of distance-adjusted means of 2.79 magnitudes indicated that mitigated satellites are more than 10 times fainter (Figure 7).

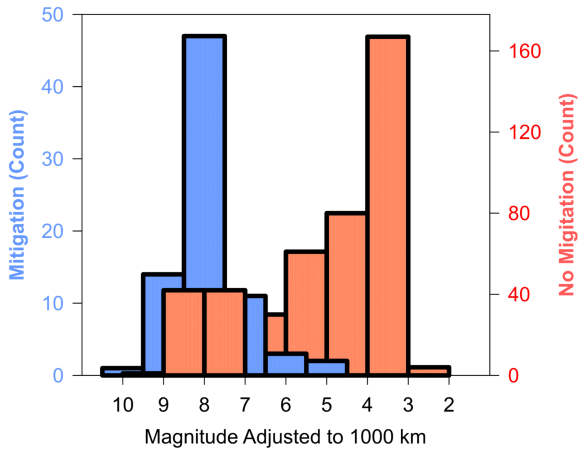

*Fig. 7. The distribution of distance-adjusted luminosity for satellites with and without brightness mitigation. Illustration from Mallama et al. (2023).*

More recently the authors have concentrated their observations on Mini satellites at small and large phase angles. These magnitudes were needed in order to fully parameterize the BRDF model discussed in Section 6. The phase function in Figure 8 demonstrates that Minis are bright at small angles and even brighter at large angles relative to mid-range angles.

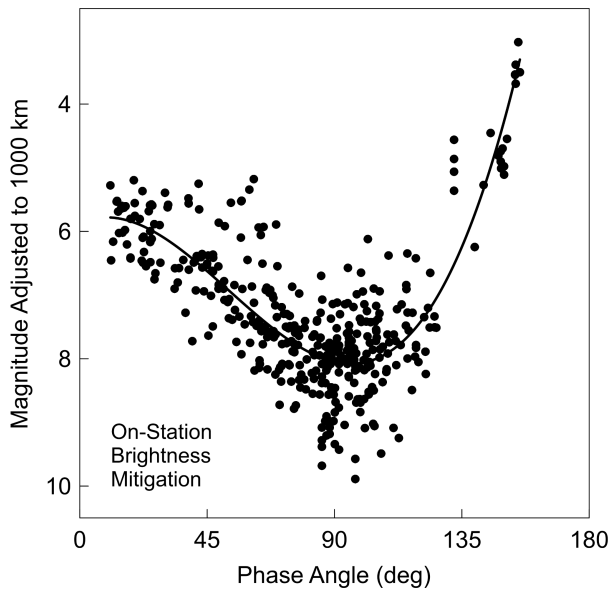

*Fig. 8. The phase function for Mini satellites illustrates their brightness as a function of angle.*

On 2023 July 14 SpaceX informed our team that they were experimenting with off-pointing the solar arrays during orbit-raising for additional brightness mitigation of the Mini satellites. So, we now distinguish between on-station and orbit-raising mitigation as well as 'no mitigation'. The magnitude distribution between these three modes is shown in Figure 9. The unmitigated satellites are brightest by far, while the luminosities of on-station and orbit-raising spacecraft are much reduced.

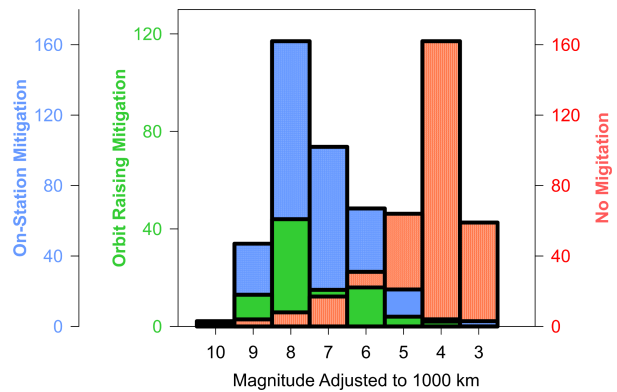

*Fig. 9. The distribution of magnitudes for on-station and orbit-raising modes as well as for no mitigation.*

*5.6 Temporal variations of brightness*

Each model of Starlink spacecraft has its own brightness characteristics as noted above. However, Cole (2021) pointed out that those of VisorSat had changed between the winter and summer of 2021. He obtained visual magnitudes in June of that year at a Sun-satellite-Earth geometry where satellites can be quite luminous. Cole found that spacecraft in that area of the sky were brighter than they had been several months earlier. This finding is discussed in terms of his BRDF model in the next Section.

This same effect was noted by Mallama (2021b) from MMT9 observations. The phase function at angles below 60°, in Figure 10, is steeper and magnitudes are brighter in the April-June 2021 period as compared to September-December 2020.

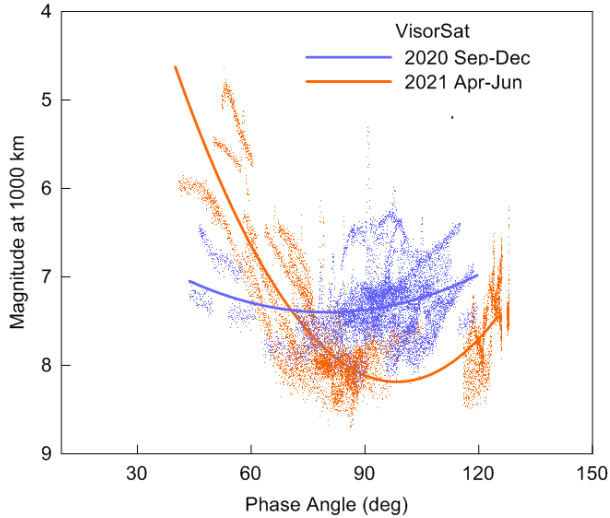

Fig. 10. The phase functions for observations recorded about six months apart. Satellites were brighter in 2021 at phase angles less than 60º and fainter at large angles. Illustration adapted from Mallama (2021b).

## 6. Brightness modeling

The physical design of a spacecraft along with the reflective properties of its surfaces account for its luminosity. Observed magnitudes or laboratory measurements may be used to parameterize a brightness model. That numerical representation can then be used to predict spacecraft luminosities for any geometry involving the satellite, the Sun and the observer. So, the spacecraft brightness model is an important tool for observation planning purposes.

Cole (2020, 2021) developed a BRDF model for VisorSat which takes account of its antenna panel and solar array. These components were in the attitude that SpaceX called shark-fin where the panel faced the Earth and the array faced the Sun as shown in Figure 11.

Cole's model considers eight angles and other factors relative to the spacecraft, the observer and the Sun. Examples are the off-base view angle measured at the spacecraft between nadir and the direction to the observer, the Sun depression angle taken between the horizontal at the satellite and the direction to the Sun, and the range measured between the spacecraft and the observer.

The model has 10 adjustable parameters such as diffuse and specular reflectivity of the antenna panel, and diffuse reflectivity of the solar array. The single output parameter is the modeled apparent magnitude.

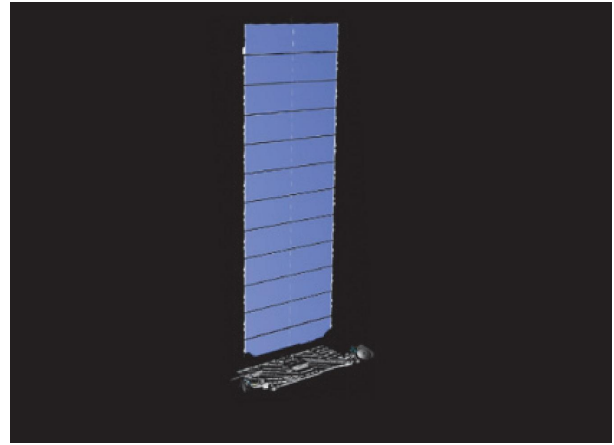

Fig. 11. Shark-fin configuration as illustrated by SpaceX.

This VisorSat model was fit to 131 magnitude records for 66 satellites at their on-station heights. Visual observations were made by the authors of the paper, and V-band measurements were obtained from the MMT9 database as well as those reported in Walker et al. (2021). The RMS residual of the model was 0.4 magnitude which Cole considered to be reasonable given the accuracy of the observations. Figure 12 illustrates the correlation between model and observed luminosity over a span of 4 magnitudes.

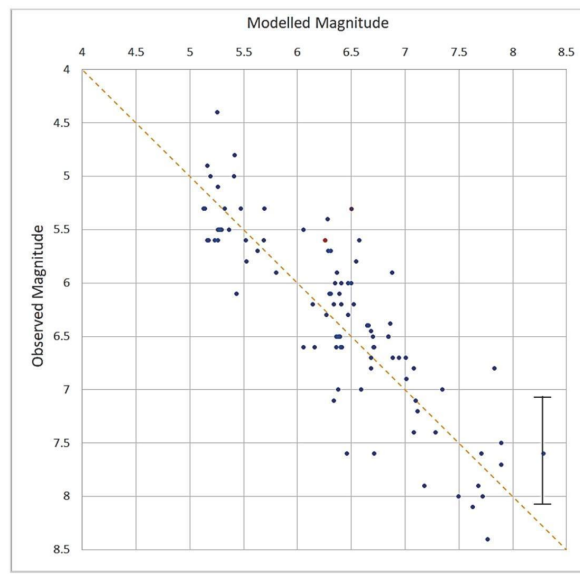

Fig. 12. The model is too faint above the dotted line and too bright below it. Illustration from Cole (2021).

Several insights were gleaned from the model. For example, the solar elevation at the observer is an

Page 8 of 12

important determinant of satellite brightness with larger negative elevations leading to greater brightness. Furthermore, maps of spacecraft brightness across the sky revealed that the satellites are generally fainter when seen nearer the horizon except for those in the anti-solar direction as shown in Figure 13.

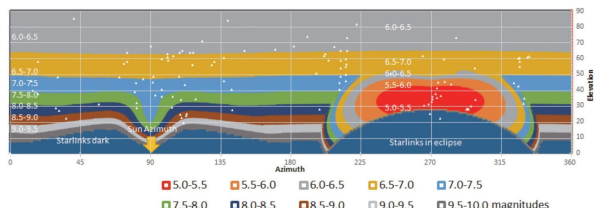

*Fig. 13. VisorSat brightness mapped onto the sky in Cartesian coordinates. The Sun is 15º below azimuth 90º. The observations are shown by small white symbols. Note the bright patch centered at azimuth 270º and elevation 35º. Illustration from Cole (2021).*

Cole also found that satellites opposite the Sun were brighter in the summer of 2021 than during the previous winter with magnitudes around 4 to 4.5 as mentioned in Section 5. The BRDF model was modified to fit these observations by changing the tilt angle of the solar array.

Fankhauser et al. (2023) modeled Starlink laser communication satellites which we refer to herein as Post-VisorSats. The physical model only required the antenna panel and the solar array. SpaceX provided BRDF functions measured in the lab for these two components. In a separate solution they constrained the BRDF parameters using magnitudes recorded by Pomenis. These models were compared to a diffuse sphere model. Both the lab BRDF and the magnitude-constrained BRDF provided a better fit to observations than the diffuse sphere model as shown in Figure 14.

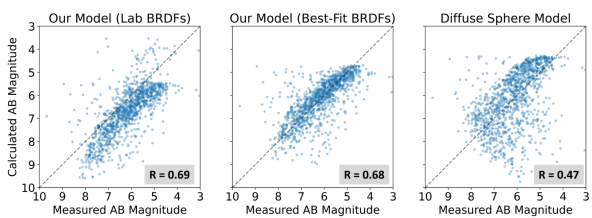

*Fig. 14. The laboratory and observation-constrained BRDF models correlate more strongly with measured magnitudes than does a diffuse sphere model. Illustration from Fankhauser et al. (2023).*

The numerical model developed by Fankhauser et al. is the first to include light reflected from the Earth to the satellite. They found that this light source causes noticeable brightness at low elevations in the solar direction as shown in Figure 15. They also point out that this excess luminosity may interfere with searches for potentially hazardous asteroids conducted during evening and morning twilight.

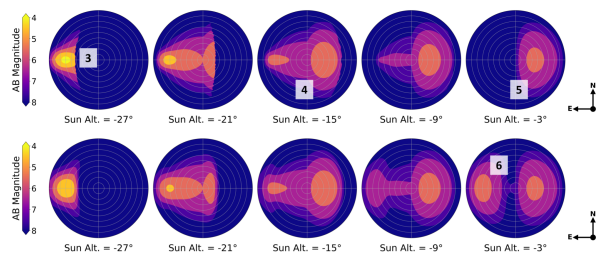

*Fig. 15. Satellite brightness derived from the laboratory BRDF model mapped onto the sky in polar coordinates. The Sun is toward the east at the elevations indicated below each map. The top row does not include light reflected from the Earth while the bottom row does. Notice the additional component of brightness in map 6 that does not appear in map 5. This extra satellite illumination comes from the Earth's day side. Illustration from Fankhauser et al. (2023).*

## 7. Discussion

This Section begins by addressing Starlink brightness in the context of observational astronomy. Then several reports concerning the impact of these satellites on specific instruments and facilities including radio telescopes are discussed. Next, an approach to mitigating satellite interference by scheduling observations to avoid them is described. Finally, the international effort aimed at protecting dark skies from bright satellite constellations is summarized.

### 7.1 Impact of Starlink on astronomy

Tyson et al. (2020) established that streaks from satellites of magnitude 7 and fainter could be successfully removed from images obtained at the Rubin Observatory. This is an important criterion since their Legacy Survey of Space and Time (LSST) is highly vulnerable to satellite interference. The magnitude 7 limit of Tyson et al. is for the g-band of the Sloan photometric system but that for the V-band is generally taken to be about the same. Both apply to satellites near the 550 km height of many Starlink spacecraft. Meanwhile, amateur astronomers refer to magnitude 6 as the limit for satellite interference because fainter objects cannot usually be seen with the unaided eye. SpaceX has stated that it aims to make on-station Starlink satellites invisible to the unaided eye. Original design Starlink satellites largely fail to meet the magnitude criteria for LSST and the unaided eye, while more VisorSats, Post-VisorSats and Minis do attain it. The surface area for full-sized Gen 2 satellites will be more than 10 times that of Gen 1 spacecraft. So, they will present a greater challenge for brightness mitigation.

*7.2 The impact on specific instruments and facilities*

Bassa et al. (2022) evaluate the impact of Starlink satellites on a variety of astronomical instruments including narrow and wide-field imagers along with long-slit and fiber-fed spectrographs. Their results indicate that the wide-field imagers were most seriously affected. They also examined observation scheduling as a mitigation strategy. Mroz et al. (2022) addressed the impact of Starlink satellites on survey observations at the Zwicky Transient Facility. They noted a large increase in the percentage of streaked images between 2019 and 2021 but concluded that their observations were not yet strongly affected. Williams et al. (2021) reported on the potential impact of Starlink satellites on the ESO optical telescopes located at Paranal and La Silla in Chile. They found the interference to be manageable at that time. They also addressed the effect of satellite noise on the ALMA radio astronomy facility at Llano de Chajnantor and reported that only one band was affected. Di Vruno et al. (2023) reported on radio noise from Starlink satellites recorded at the LOFAR radio telescope. They detected interference at frequencies between 110 and 188 MHz. The authors characterise this noise as 'unintended' and point out that it is not subject to existing regulations.

*7.3 The scheduling approach*

Hu et al. (2022) examine the effectiveness of adjusting the scheduler algorithm for the LSST to avoid satellites at the cost of decreased efficiency in executing other science goals. They find that the need for this mitigation strategy will depend on the overall impact of satellite streaks. They further state the impact is not yet well known due to incomplete information about satellite luminosities. That knowledge is incomplete, as they said, but it is rapidly growing.

*7.4 Protecting dark skies*

The observations, analyses and models described in this paper quantify satellite brightness. This research contributes to a larger effort aimed at mitigating the adverse effects of spacecraft on astronomy. We have summarized our own research and that of numerous other investigators, but the complete body of literature on satellite interference is too extensive to include here. Many more useful papers can be found in the proceedings of Dark and Quiet Skies conferences (Walker et al. 2021 and Walker and Benvenuti 2022).

The International Astronomical Union established the [Centre for the Protection of Dark and Quiet Skies from Satellite Constellation Interference](#) in 2022. This organization coordinates world-wide efforts aimed at mitigating the negative impact of satellite constellations. The CPS has 'hubs' that specialize in public policy, industry and technology, and community engagement. Their SatHub offers an astronomical data repository, an orbital solutions portal, software tools, a training curriculum and real-time collaboration.

**8. Conclusions**

The Original design of Starlink satellites concerned astronomers because their large number and great brightness were seen as a serious threat to celestial observations. SpaceX responded to these concerns by changing the physical design of their VisorSat and Post-VisorSat models and by modifying their conops for spacecraft in orbit. Meanwhile photometric observers verified that these alterations substantially mitigated brightness.

There were new concerns when SpaceX revealed that their second generation satellites would be larger. The most recent observations indicate that the first model of Gen 2 spacecraft, called Mini, is actually dimmer than those of Gen 1. The full-sized satellites to come later will present a greater challenge to the company's brightness mitigation efforts. Future observations will measure the brightness of those very large spacecraft and monitor the luminosity of earlier models.